\documentclass[11pt]{article}
\usepackage{amsmath}
\usepackage{amssymb}
\usepackage{graphicx}
\usepackage{color}
\usepackage{orcidlink}
\usepackage{hyperref}
\hypersetup{colorlinks=true, linkcolor=blue, citecolor=blue, urlcolor=blue}
\usepackage{caption}
\usepackage{subcaption}
\begin{document}
\baselineskip=16pt

\begin{center}
{\large {\bf Thermal Behavior of Generalized Black-Bounce Black Hole Model}}
\end{center}

\vspace{0.3cm}

\begin{center}

{\bf Allan. R. P. Moreira\orcidlink{0000-0002-6535-493X}}\footnote{\bf allan.moreira@fisica.ufc.br }\\
\vspace{0.1cm}
{\it Reserach Center for Quantum Physics, Huzhou University, Huzhou, 313000, P. R. China}\\
{\it Secretaria da Educa\c{c}\~{a}o do Cear\'{a} (SEDUC), Coordenadoria Regional de Desenvolvimento da Educa\c{c}\~{a}o (CREDE 9),  Horizonte, Cear\'{a}, 62880-384, Brazil}\\
\vspace{0.1cm}
{\bf Abdelmalek Bouzenada\orcidlink{0000-0002-3363-980X}}\footnote{\textbf{abdelmalekbouzenada@gmail.com (Corresp. author) }}\\
\vspace{0.1cm}
{\it Laboratory of Theoretical and Applied Physics, Echahid Cheikh Larbi Tebessi University 12001, Algeria}\\
\vspace{0.1cm}
{\bf Shi-Hai Dong\orcidlink{0000-0002-0769-635X}}\footnote{\bf dongsh2@yahoo.com }\\
\vspace{0.1cm}
{\it Reserach Center for Quantum Physics, Huzhou University, Huzhou, 313000, P. R. China}\\
{\it Centro de Investigaci\'{o}n en Computaci\'{o}n, Instituto Polit\'{e}cnico Nacional, UPALM, CDMX 07700, Mexico.}\\
\vspace{0.1cm}
{\bf Guo-Hua Sun\orcidlink{0000-0002-0689-2754}}\footnote{\bf sunghdb@yahoo.com }\\
\vspace{0.1cm}
{\it Centro de Investigaci\'{o}n en Computaci\'{o}n, Instituto Polit\'{e}cnico Nacional, UPALM, CDMX 07700, Mexico.}\\
\vspace{0.1cm}
{\bf Faizuddin Ahmed\orcidlink{0000-0003-2196-9622}}\footnote{\bf faizuddinahmed15@gmail.com }\\
\vspace{0.1cm}
{\it Department of Physics, The Assam Royal Global University, Guwahati, 781035, Assam, India}\\
\vspace{0.35cm}
\end{center}

\vspace{0.2cm}
\begin{abstract}

In this work, we tested the thermal behavior of a class of regular black hole solutions defined as generalized black-bounce spacetimes. We introduce several novel configurations governed by different mass functions and geometric deformations, illustrated by parameters controlling regularity and horizon structure. Using the Hamilton–Jacobi tunneling method, we compute the Hawking temperature associated with each model and analyze its dependence on the underlying parameters. We find that all proposed geometries are free of curvature singularities and exhibit positive, well-defined quasi-local masses in the Hernandez–Misner–Sharp formalism. Also, we demonstrate that these models may possess multiple horizons, including extremal and asymmetric cases, while typically violating classical energy conditions in the vicinity of the bounce. Our results show and illustrate the structure and thermodynamic stability of these regular solutions.
\end{abstract}

\vspace{0.1cm}

\textbf{Keywords}:{ Hawking Temperature; Black Holes; Black-Bounce Spacetimes; Horizon Structure; Local Mass }

\vspace{0.1cm}

\textbf{PACS:} {04.70.-s,04.50.Kd,11.30.Cp,04.60.-m }

\section{Introduction}

It is fascinating to explore the diverse landscape of regular black holes and their connections to ``black bounce'' solutions, particularly those featuring a minimal areal radius in the T-region, where the radial coordinate becomes timelike, or situated on a horizon, as discussed in previous studies \cite{A1,A2,A3,A4}. Also, the spacetimes tested in \cite{A5} exhibit a de Sitter late-time asymptotic, positioning them as potential candidates for viable cosmological models. In this case, further enriching this discussion, \cite{A6} illustrates regular solutions incorporating a phantom scalar field and an electromagnetic field, yielding a variety of global structures, including configurations with up to four horizons. Also, the stability of these solutions was rigorously analyzed in \cite{A7}, revealing that most configurations are unstable under spherically symmetric perturbations, with the exception of a unique class of black universes where the event horizon aligns with the minimal area function. Beyond black holes, another intriguing class of solutions arises in the form of wormholes \cite{A8}, which are horizon-free and regular. In another study, the Ellis-Bronnikov wormhole, which emerged in 1973 \cite{A9,A10}, illustrated by numerous other solutions \cite{A11,A12,A13,A14,A15,A16}, all properties of a bridging structure known as a throat, typically located at the radial coordinate’s center. A particularly innovative solution merging wormhole and regular black hole features is the Simpson-Visser spacetime \cite{A17}, where a tunable parameter allows the solution to transition between a regular black hole with a throat and a pure wormhole, reverting to the Schwarzschild solution when the parameter vanishes. This model has been extensively tested \cite{A18,A9,A20,A21,A22}, alongside other suggested solutions \cite{A23,A24,A25,A26,A27,A28,A29,A30,A31}, including those incorporating a cloud of strings \cite{A32,A33}. Intriguingly, such solutions cannot be described solely by nonlinear electrodynamics or a scalar field but instead require a coupling between nonlinear electrodynamics and a phantom scalar field, with the bounce parameter corresponding to the magnetic charge parameters \cite{A34,A35,A36,A37}. Also, another study shows and tests that black bounce solutions can also emerge from electric sources \cite{A30}-\cite{A37}, further expanding the theoretical possibilities in this field.

In the study of black hole (BH) properties \cite{B1,B2,B3,B4,B5,B6,B7,B8,B9,B10}, S. Hawking was the first to explain that BHs can spontaneously emit particles at a temperature inversely related to their mass \cite{Hawking1974}, a discovery that significantly influenced both classical and quantum gravity theories \cite{Miao2017}. Other studies have extensively tested the Hawking radiation spectrum and temperature, aiming to uncover potential insights into quantum gravity \cite{Page1975, Page1976, Parikh2000}. Among the key contributions in this field, Page calculated particle emission rates for both nonrotating and rotating BHs in Refs. \cite{Page1975, Page1976}. A pivotal advancement came in 1999 when Wilczek and Parikh \cite{Parikh2000} proposed that Hawking radiation could be interpreted as a quantum tunneling phenomenon, wherein particles traverse the BH's contracting horizon. These results have also been extended to various BH models, including Einstein–Gauss–Bonnet de Sitter BHs \cite{Zhang2018} and charged BHs ($Q$) \cite{Miao2017, Chowdhury2020}. In this context, Bardeen, Carter, and Hawking \cite{Bardeen1973} formulated the laws of BH thermodynamics, illustrating the relation between surface gravity and temperature, as well as between horizon area and entropy. Furthermore, more studies have revealed non-thermal radiation emissions lacking a well-defined temperature in extremal Reissner–Nordström and Kerr BHs \cite{Good2020, Good2021}, further enriching our illustration of BH dynamics.

Our study tests a broad class of generalized black-bounce black hole geometries by introducing new mass function models and showing their effects on the spacetime structure and temperature behavior. By varying key parameters, including the bounce parameter $a$, the mass deformation indices $n$ and $k$, and the scale parameter $r_0$, we construct regular black hole solutions that interpolate between classical Schwarzschild and wormhole-like configurations. We analyze the behavior of the metric function $f(r)$ and the Hernandez–Misner–Sharp quasi-local mass $M_{\text{HMS}}(r)$ for each model, confirming that the geometries remain regular, horizon-forming, and possess positive mass distributions throughout the manifold. Furthermore, we calculate the Hawking temperature $T_{BH}$ using the Hamilton–Jacobi tunneling method and demonstrate that the temperature decreases with increasing $a$ and $n$, indicating suppressed radiation and enhanced thermodynamic stability. These results show the role of bounce-driven regularization in constructing physically admissible black hole models of thermal influence.

The structure of this article is organized as follows: in Section~\ref{sec1}, we introduce the framework of generalized black-bounce spacetimes, highlighting the relevant geometric features and the mass functions that define the models. In Section~\ref{sec2}, we analyze specific cases of these spacetimes by exploring different parametrizations that regularize the central region. Section III is devoted to computing the Hawking temperature using the tunneling method based on the Hamilton–Jacobi formalism, emphasizing how the model parameters influence the thermal behavior of the black hole. Finally, in Section~\ref{sec00}, we present our conclusions, summarizing the key results and outlining potential directions for future investigations.

\section{Generalized black-bounce spacetimes}\label{sec1}

A general static and spherically symmetric spacetime can locally be expressed using the following line element:
\begin{eqnarray}
ds^2 = f(r)\,dt^2 - \frac{dr^2}{f(r)} - \Sigma^2(r)\left(d\theta^2 + \sin^2\theta\, d\phi^2\right)\,,\label{ele}
\end{eqnarray}
where the metric functions $ f(r) $ and $ \Sigma(r) $ are arbitrary and determined by the gravitational field equations and matter content. The locations of potential horizons correspond to the roots of $ f(r) = 0 $, and the determinant of the metric tensor is $ g = -\Sigma^4(r) \sin^2\theta $. The areal radius is given by $ A(r) = 4\pi \Sigma^2(r) $. This coordinate system is commonly referred to as the Buchdahl form~\cite{finch-skea,petarpa1,petarpa2,Semiz:2020}.

The Einstein equations read:
\begin{eqnarray}
R_{\mu\nu} - \frac{1}{2}g_{\mu\nu}R = \kappa^2 T_{\mu\nu}\,,\label{eqEinstein}
\end{eqnarray}
with $ R_{\mu\nu} $ the Ricci tensor, $ R $ the Ricci scalar, and $ T_{\mu\nu} $ the energy-momentum tensor. We adopt the metric signature $(+,-,-,-)$. Given the Levi-Civita connection
\begin{eqnarray}
\Gamma^{\alpha}{}_{\mu\nu}=\frac{1}{2} g^{\alpha\beta}\left(\partial_{\mu}g_{\nu\beta}+\partial_{\nu}g_{\mu\beta}-\partial_{\beta}g_{\mu\nu}\right),
\end{eqnarray}
the Riemann tensor is defined as 
\begin{eqnarray}
R^{\alpha}{}_{\beta\mu\nu}=\partial_{\mu}\Gamma^{\alpha}{}_{\beta\nu}-\partial_{\nu}\Gamma^{\alpha}{}_{\beta\mu}+\Gamma^{\sigma}{}_{\beta\nu}\Gamma^{\alpha}{}_{\sigma\mu}-\Gamma^{\sigma}{}_{\beta\mu}\Gamma^{\alpha}{}_{\sigma\nu}.
\end{eqnarray}
Adopting geometrized units where $ G = c = 1 $, we set $ \kappa^2 = 8\pi $. For an anisotropic matter distribution, in regions where the temporal coordinate remains timelike ($ f(r) > 0 $), the mixed form of the energy-momentum tensor is:
\begin{eqnarray}
T^{\mu}{}_{\nu} = \mathrm{diag}[\rho, -p_1, -p_2, -p_2]\,,\label{EMT}
\end{eqnarray}
where $ \rho $ represents the energy density, and $ p_1 $, $ p_2 $ denote the radial and tangential pressures. By inserting the metric \eqref{ele} into \eqref{eqEinstein}, one obtains:
\begin{eqnarray}
\rho &=& -\frac{1}{\kappa^2 \Sigma^2} \left[ \Sigma f' \Sigma' + 2f \Sigma'' \Sigma + f \Sigma'^2 - 1 \right]\,,\label{density+}\\
p_1 &=& \frac{1}{\kappa^2 \Sigma^2} \left[ \Sigma f' \Sigma' + f \Sigma'^2 - 1 \right]\,,\label{p1+}\\
p_2 &=& \frac{1}{2\kappa^2 \Sigma} \left[ \Sigma f'' + 2f' \Sigma' + 2f \Sigma'' \right]\,,\label{p2+}
\end{eqnarray}
where primes denote derivatives with respect to $ r $.

Inside regions where the coordinate $ t $ becomes spacelike ($ f(r) < 0 $), the stress-energy tensor takes the form:
\begin{eqnarray}
T^{\mu}{}_{\nu} = \mathrm{diag}[-p_1, \rho, -p_2, -p_2]\,,\label{EMT2}
\end{eqnarray}
and the field equations yield:
\begin{eqnarray}
\rho &=& -\frac{1}{\kappa^2 \Sigma^2} \left[ \Sigma f' \Sigma' + f \Sigma'^2 - 1 \right]\,,\label{density-}\\
p_1 &=& \frac{1}{\kappa^2 \Sigma^2} \left[ \Sigma f' \Sigma' + 2f \Sigma'' \Sigma + f \Sigma'^2 - 1 \right]\,,\label{p1-}\\
p_2 &=& \frac{1}{2\kappa^2 \Sigma} \left[ \Sigma f'' + 2f' \Sigma' + 2f \Sigma'' \right]\,,\label{p2-}
\end{eqnarray}

At the horizon, where $ f(r) = 0 $, these expressions reduce to:
\begin{eqnarray}
\rho = -p_1 &=& -\frac{1}{\kappa^2 \Sigma^2} \left[ \Sigma f' \Sigma' - 1 \right]\,,\label{density0}\\
p_2 &=& \frac{1}{2\kappa^2 \Sigma} \left[ \Sigma f'' + 2f' \Sigma' \right]\,,\label{p20}
\end{eqnarray}
where continuity in $ \rho $ across the horizon requires $ \rho = -p_1 $, a condition well-established in the literature~\cite{dbh1,dbh2}.

The trace of the energy-momentum tensor is calculated as:
\begin{equation}
T = T^\mu{}_\mu = \rho - p_1 - 2p_2 = 
-\frac{ \Sigma^2 f'' + 4\Sigma (\Sigma'' f + \Sigma' f') + 2f \Sigma'^2 - 2 }{\kappa^2 \Sigma^2 }\,.
\end{equation}
This result holds uniformly, regardless of whether the region lies inside or outside any horizon.

To ensure the regularity of the stress-energy components and of the spacetime geometry, the following smoothness and non-degeneracy conditions must be satisfied:
\begin{itemize}
\item $ \Sigma(r) \neq 0 $ for all $ r $,
\item $ \Sigma'(r) $ and $ \Sigma''(r) $ must remain finite,
\item $ f(r) $, $ f'(r) $, and $ f''(r) $ must also remain finite.
\end{itemize}

A natural definition of a quasi-local mass function, often attributed to Hernandez, Misner, and Sharp~\cite{Hernandez:1966, Misner:1964, Maeda:2007, Nielsen:2008, Abreu:2010, Faraoni:2020}, arises from analyzing the curvature component:
\begin{eqnarray}
R^{\theta\phi}{}_{\theta\phi} = -\frac{2M_{\text{HMS}}(r)}{\Sigma^3(r)} = \frac{f(r) \Sigma'^2(r) - 1}{\Sigma^2(r)}\,,\label{Rmass}
\end{eqnarray}
leading to the expression:
\begin{eqnarray}
M_{\text{HMS}}(r) = \frac{1}{2}\Sigma(r) \left[ 1 - f(r)\Sigma'^2(r) \right]\,.\label{mass}
\end{eqnarray}
Inverting this, one can write:
\begin{eqnarray}
f(r) = \frac{1 - 2M_{\text{HMS}}(r)/\Sigma(r)}{\Sigma'^2(r)}\,,\label{f}
\end{eqnarray}
or, alternatively, adopt a reparametrization:
\begin{eqnarray}
f(r) = 1 - \frac{2M(r)}{\Sigma(r)}\,,\label{fm}
\end{eqnarray}
where $ M(r) $ is now interpreted as an effective mass function embedded in the metric ansatz, without necessarily preserving its quasi-local mass interpretation.

\section{Specific black-bounce spacetimes}\label{sec2}

Let us now examine a broad family of black-bounce geometries that extend the Simpson-Visser spacetime by introducing more general functional forms for $\Sigma(r)$, $M(r)$, and $f(r)$. In this framework, we define:
\begin{equation}
\Sigma(r) = \sqrt{r^2 + a^2}, \qquad
M(r) = \frac{m \, \Sigma(r) \, r^k}{\left(r^{2n} + a^{2n}\right)^{(k+1)/(2n)}}, \qquad
f(r) = 1 - \frac{2M(r)}{\Sigma(r)}\label{newBBS}.
\end{equation}

Here, $n$ and $k$ are arbitrary positive integers that control the asymptotic behavior and regularity properties of the solution. This construction is influenced by the mass profile proposed by Fan and Wang~\cite{Fan.Wang} in the context of regular black holes. When choosing $n=1$ and $k=0$, the above expressions reduce precisely to the Simpson-Visser metric
\begin{eqnarray}
ds^2 = \Bigg(1 - \frac{2m}{\sqrt{r^2 + a^2}}\Bigg)dt^2 - \frac{dr^2}{\Big(1 - \frac{2m}{\sqrt{r^2 + a^2}}\Big)} - (r^2 + a^2)\left(d\theta^2 + \sin^2\theta\, d\phi^2\right).
\end{eqnarray}
Moreover, taking the limit $a \rightarrow 0$ recovers the classical Schwarzschild geometry, regardless of the values of $n$ and $k$.

It is worth noting that, due to the specific presence of the $a^2$ term in $\Sigma(r)$, this model does not encompass some well-known regular black hole solutions such as those proposed by Bardeen, Hayward, or Frolov. Nonetheless, the formulation allows for the construction of a wide range of novel black-bounce configurations, several of which we shall analyze in detail in the following sections.

\begin{figure}[ht]
\centering
\includegraphics[height=5cm]{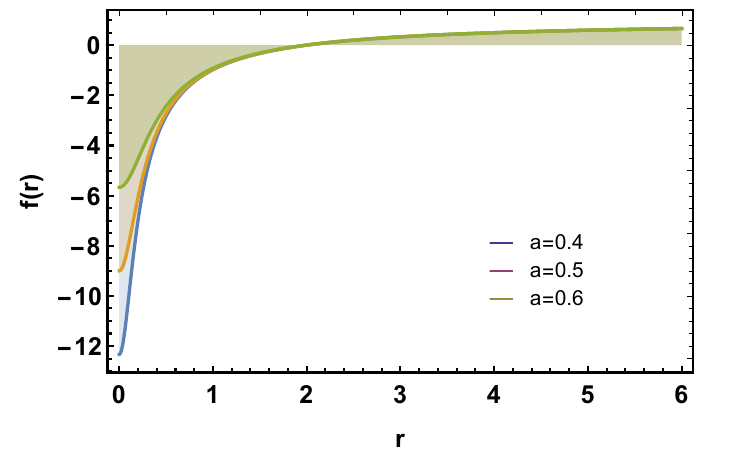}\\
\caption{Metric function $f(r)$ with $m=1$.}
\label{fig1}
\end{figure}

This figure (\ref{fig1}) shows the metric function $f(r)$ for the generalized black-bounce spacetime using the Simpson–Visser profile, with the fixed mass parameter $m = 1$ and varying bounce parameters $a = 0.4, 0.5, 0.6$. As $a$ increases, the minimum of the function $f(r)$ becomes shallower, indicating that increasing the bounce parameter softens the central gravitational potential and horizons.

\subsection{Model $n=1$ and $k=2$}

Consider now the particular case of the generalized black-bounce model described by equation~\eqref{newBBS}, with parameters $n = 1$ and $k = 2$. In this scenario, the functions take the form:
\begin{equation}
\Sigma(r) = \sqrt{r^2 + a^2}, \qquad f(r) = 1 - \frac{2m r^2}{(r^2 + a^2)^{3/2}} \label{mod2}.
\end{equation}
The metric function $f(r)$ coincides formally with that of the Bardeen regular black hole when identifying $a \leftrightarrow q$. However, the underlying geometry differs significantly from the Bardeen case, as the function $\Sigma(r)$, which determines the spatial geometry, explicitly includes $a^2$, modifying the overall structure. Notably, the Bardeen configuration would require $a \to 0$ in $\Sigma(r)$ while retaining $a \neq 0$ in $f(r)$, something not applicable here.

\begin{figure}[ht]
\centering
\includegraphics[height=5cm]{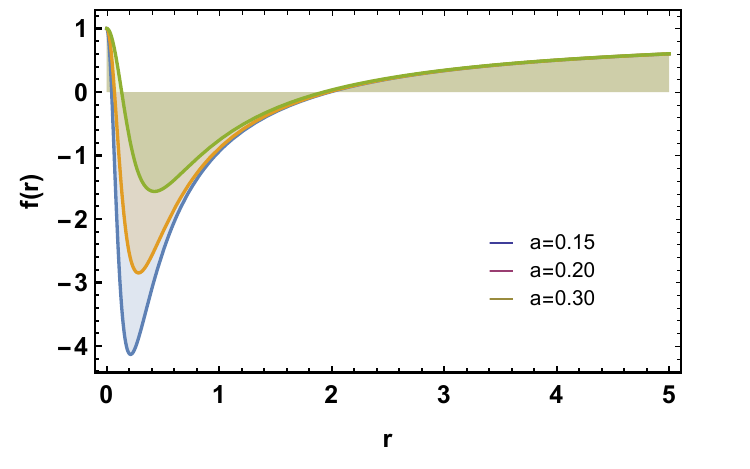}\\
\caption{Metric function $f(r)$ with $m=1$.}
\label{fig2}
\end{figure}

Here (Figure (\ref{fig2})), the metric function $f(r)$ is plotted for the specific model with parameters $n = 1$ and $k = 2$, and fixed mass $m = 1$. The bounce parameter takes values $a = 0.15, 0.20, 0.30$. The plot shows that as $a$ increases, the potential well becomes less deep and the position of the horizons shifts, showing how the geometry evolves with the bounce parameter in this modified Bardeen-like model.

The corresponding Hernandez-Misner-Sharp quasi-local mass, derived from equation~\eqref{mass}, becomes:
\begin{equation}
M_{\text{HMS}}(r) = \frac{a^2}{2\sqrt{r^2 + a^2}} + \frac{m r^4}{(r^2 + a^2)^2}.
\end{equation}
This expression ensures that the mass is strictly non-negative across all $r$. Furthermore, the mass satisfies the limiting behavior:
\begin{equation}
\lim_{r \to 0} M_{\text{HMS}}(r) = \frac{a}{2}, \qquad \lim_{r \to \infty} M_{\text{HMS}}(r) = m.
\end{equation}

\begin{figure}[ht]
\centering
\includegraphics[height=5cm]{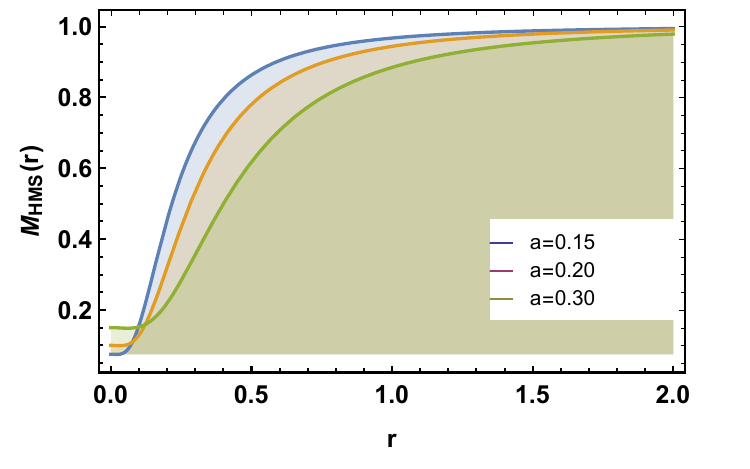}\\
\caption{Hernandez–Misner–Sharp quasi-local mass with $m=1$.}
\label{fig22}
\end{figure}

This figure (\ref{fig22}) displays the Hernandez–Misner–Sharp quasi-local mass $M_{\text{HMS}}(r)$ for the model with $n = 1$, $k = 2$, and $m = 1$. The curves are shown for $a = 0.15, 0.20, 0.30$. The mass function is strictly positive and increases smoothly from $a/2$ at the origin to $m$ at spatial infinity, showing how the bounce parameter $a$ regulates the near-core behavior of mass.

One may also explore a broader class of models by maintaining $k = 2$ while varying the value of the integer $n$. In such cases, the resulting geometries preserve similar structural features to the $n = 1$ case discussed above. However, it can be verified that the energy density evaluated outside the outermost horizons consistently takes on negative values, a point that will be discussed further in the next sections.

\subsection{Model $n=2$ and $k=0$}\label{SS:n=2+k=0}

Let us now examine the case defined by setting $n = 2$ and $k = 0$ in equation~\eqref{newBBS}. In this configuration, the defining functions take the form:
\begin{equation}
\Sigma(r) = \sqrt{r^2 + a^2}, \qquad f(r) = 1 - \frac{2m}{(r^4 + a^4)^{1/4}} \label{mod1}.
\end{equation}

\begin{figure}[ht]
\centering
\includegraphics[height=5cm]{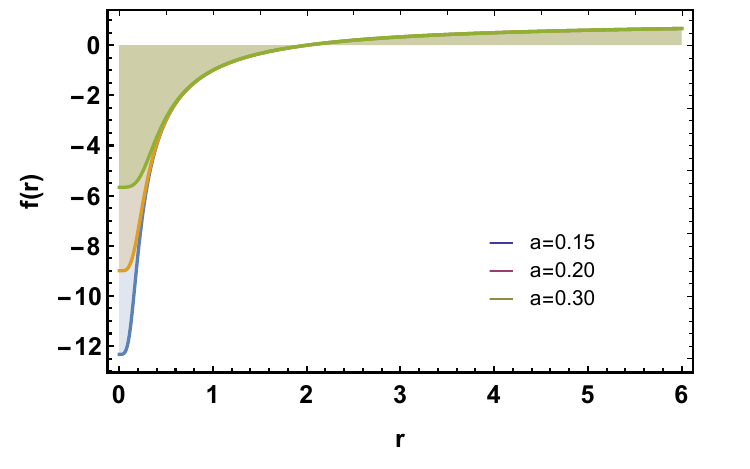}\\
\caption{Metric function $f(r)$ with $m=1$.}
\label{fig3}
\end{figure}

In this figure (\ref{fig3}), the metric function $f(r)$ is shown for the case $n = 2$, $k = 0$, and $m = 1$, with bounce parameter values $a = 0.15, 0.20, 0.30$. Also, the plots illustrate a distinct shape from the earlier model, showing how different choices of $n$ and $k$ affect the curvature structure and potential location of horizons in the generalized geometry.

The associated Hernandez-Misner-Sharp mass function, using expression~\eqref{mass}, becomes:
\begin{equation}
M_{\text{HMS}}(r) = \frac{a^2}{2\sqrt{r^2 + a^2}} + \frac{m r^2}{\sqrt{r^2 + a^2}\, (r^4 + a^4)^{1/4}}.
\end{equation}
This function remains positive for all values of $r$, with the asymptotic behavior given by:
\begin{equation}
\lim_{r \to 0} M_{\text{HMS}}(r) = \frac{a}{2}, \qquad \lim_{r \to \infty} M_{\text{HMS}}(r) = m.
\end{equation}

\begin{figure}[ht]
\centering
\includegraphics[height=5cm]{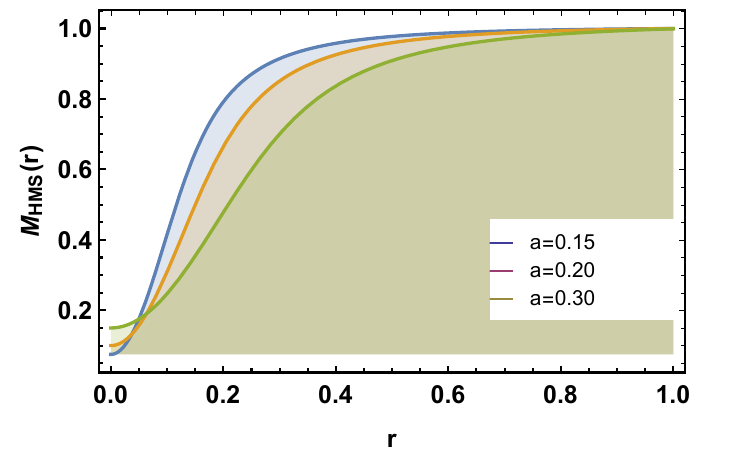}\\
\caption{Hernandez–Misner–Sharp quasi-local mass with $m=1$.}
\label{fig33}
\end{figure}

This plot (Figure (\ref{fig33})) presents the quasi-local mass $M_{\text{HMS}}(r)$ for the $n = 2$, $k = 0$ model with $m = 1$, and bounce parameters $a = 0.15, 0.20, 0.30$. Also, the behavior again confirms positivity and smoothness, with mass ranging from $a/2$ at $r = 0$ to asymptotically approaching $m$, confirming a regular, horizon-respecting profile.

When considering a broader family of solutions by varying the integer parameter $n$ while keeping $k = 0$ fixed, one finds that the qualitative features of the resulting geometries are largely preserved and closely resemble those of the original Simpson-Visser black-bounce construction. This observation suggests that further modifications in the behavior of the spacetime geometry are more effectively achieved by altering the value of $k$, which we shall explore in the subsequent analysis.

\subsection{Model $M(r)=m\cos^{2n}\left[r_0/\Sigma(r)\right]$}

Let us now explore an alternative construction for the mass function by defining
\begin{equation}
M(r) = m \cos^{2n} \left( \frac{r_0}{\Sigma(r)} \right),
\end{equation}
where the function $\Sigma(r)$ is retained as $\Sigma(r) = \sqrt{r^2 + a^2}$. This leads to the corresponding lapse function:
\begin{equation}
f(r) = 1 - \frac{2 M(r)}{\Sigma(r)} = 1 - \frac{2m \cos^{2n}\left( r_0 / \Sigma \right)}{\Sigma} \label{mod3}.
\end{equation}

\begin{figure}[ht]
\centering
\includegraphics[height=4.5cm]{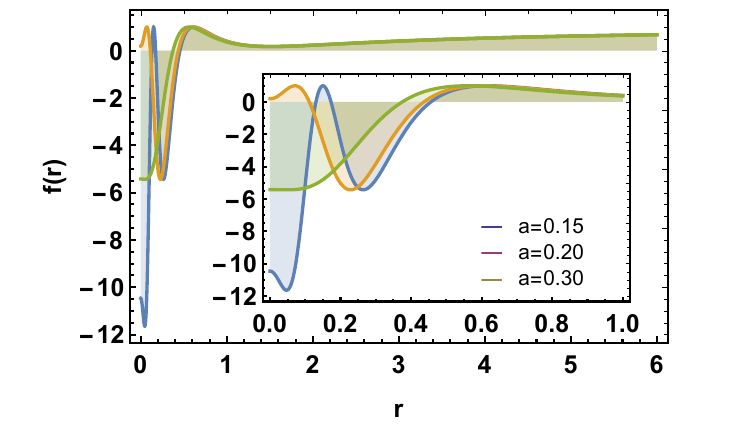}
\includegraphics[height=4.5cm]{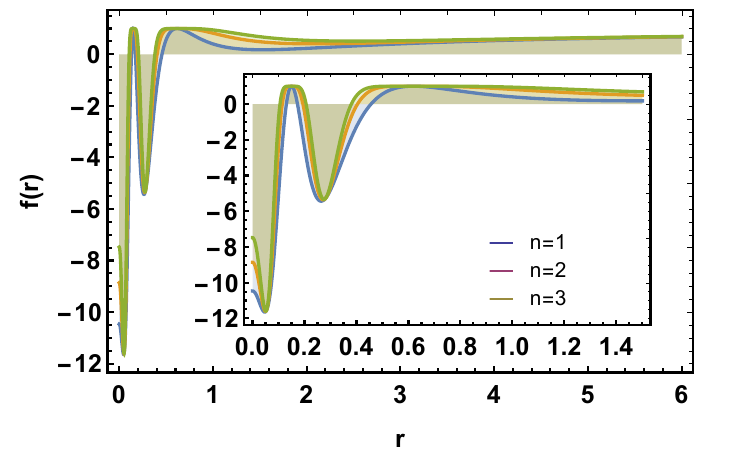}\\
\caption{Metric function $f(r)$ with $m=1$. (a) $n=1$. (b) $a=0.15$.}
\label{fig4}
\end{figure}

These plots (a) and (b) in Figure (\ref{fig4}) show the metric function $f(r)$ for the cosine-dependent mass model $M(r) = m \cos^{2n}(r_0/\Sigma)$, with fixed $m = 1$. In  (a) : $n = 1$, and bounce parameters are $a = 0.15, 0.20, 0.30$.   In  (b) : $a = 0.15$, with $n = 1, 2, 3$. Also, these figures reveal how both $n$ and $a$ control the regularity and structure of the lapse function, modifying the throat and horizon behavior.

This setup generalizes the Simpson-Visser geometry by introducing a cosine-dependent deformation in the mass profile. In the limit $n \to 0$, we recover the Simpson-Visser model, since $\cos^0(x) = 1$ and thus $M(r) = m$, while still preserving the modified areal radius $\Sigma(r) = \sqrt{r^2 + a^2}$.

In the asymptotic region, as $r \to \infty$, the geometry behaves like Schwarzschild:
\begin{equation}
\lim_{r \to \infty} f(r) = 1,
\end{equation}
and for large $r$, the expansion yields the familiar form $f(r) \sim 1 - 2m / r$.

At the origin, the behavior of the lapse function is governed by
\begin{equation}
\lim_{r \to 0} f(r) = 1 - \frac{2m \cos^{2n} \left( r_0 / a \right)}{a},
\end{equation}
showing that the central region is regular provided $a \ne 0$.

The associated Hernandez-Misner-Sharp mass function for this configuration becomes:
\begin{equation}
M_{\text{HMS}}(r) = \frac{a^2}{2\sqrt{r^2 + a^2}} + \frac{m r^2 \cos^{2n} \left( \frac{r_0}{\sqrt{r^2 + a^2}} \right)}{r^2 + a^2}.
\end{equation}

This mass function is manifestly positive for all $r$, with the following limits:
\begin{equation}
\lim_{r \to 0} M_{\text{HMS}}(r) = \frac{a}{2}, \qquad \lim_{r \to \infty} M_{\text{HMS}}(r) = m.
\end{equation}
These properties confirm the regular character of the geometry and motivate further investigation into the structure of the horizons and energy conditions for various values of $n$ and $r_0$.

\begin{figure}[ht]
\centering
\includegraphics[height=4.5cm]{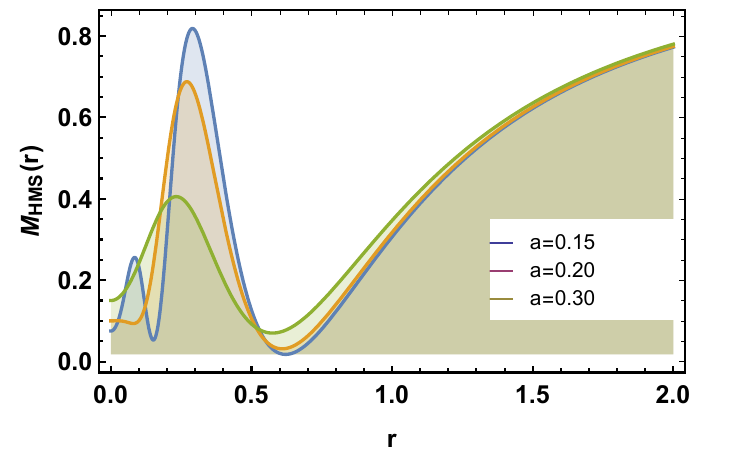}
\includegraphics[height=4.5cm]{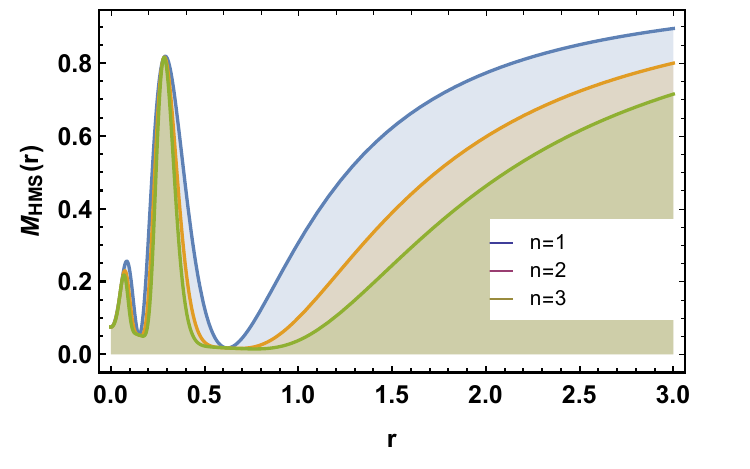}\\
\caption{Hernandez–Misner–Sharp quasi-local mass with $m=1$. (a) $n=1$. (b) $a=0.15$.}
\label{fig44}
\end{figure}

This figure (\ref{fig44}) illustrates the quasi-local mass $M_{\text{HMS}}(r)$ for the same cosine-type model with $m = 1$.(a)  varies $a = 0.15, 0.20, 0.30$ for fixed $n = 1$. (b)  varies $n = 1, 2, 3$ with fixed $a = 0.15$. In both cases, the mass grows smoothly and monotonically, show how both parameters shape the internal structure.

\subsection{Model $M(r)=m\arctan^n(r/a)\;(\Sigma/r) (2/\pi)^n$}

We now propose a modified mass profile that ensures a positive energy density across the spacetime. Let us consider the particular choice
\begin{equation}
M(r) = m \left( \frac{2}{\pi} \right)^n \arctan^n \left( \frac{r}{a} \right) \frac{\Sigma(r)}{r},
\end{equation}
which leads to the lapse function
\begin{equation}
f(r) = 1 - \frac{2 M(r)}{\Sigma(r)} = 1 - \frac{2m}{r} \left( \frac{2}{\pi} \right)^n \arctan^n \left( \frac{r}{a} \right) \label{mod4}.
\end{equation}

\begin{figure}[ht]
\centering
\includegraphics[height=4.5cm]{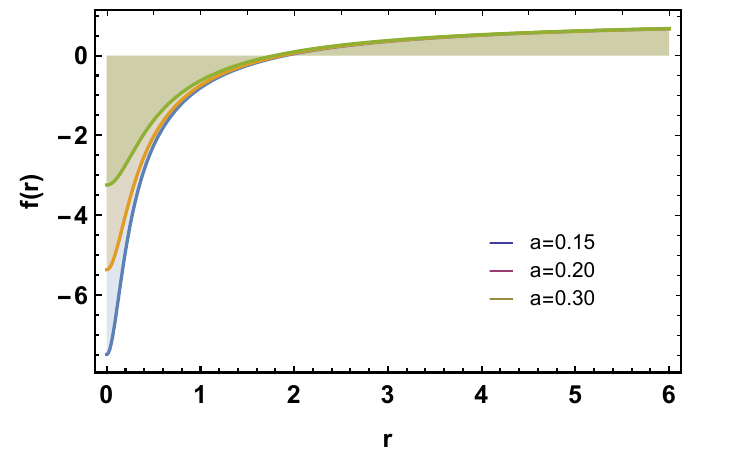}
\includegraphics[height=4.5cm]{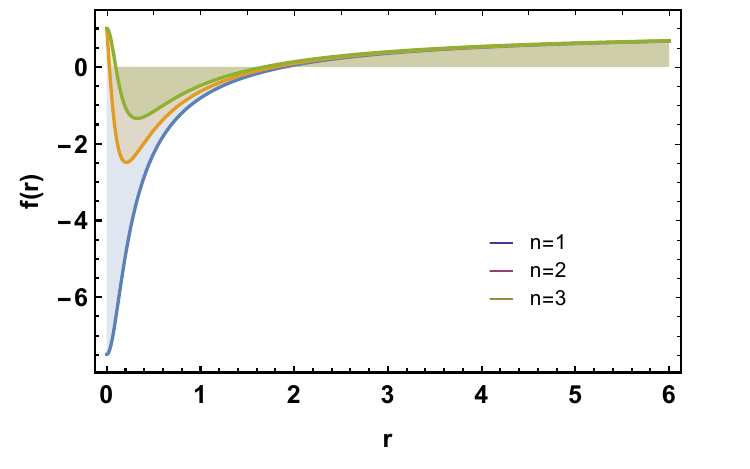}\\
\caption{Metric function $f(r)$ with $m=1$. (a) $n=1$. (b) $a=0.15$.}
\label{fig5}
\end{figure}

Here (Figure (\ref{fig5})), the metric function $f(r)$ is plotted for the model $M(r) = m \left(\frac{2}{\pi}\right)^n \arctan^n(r/a) \frac{\Sigma}{r}$, with fixed $m = 1$.(a) : $n = 1$, and $a = 0.15, 0.20, 0.30$. (b) : $a = 0.15$, and $n = 1, 2, 3$. The figures demonstrate that both higher $n$ and larger $a$ reduce the depth of $f(r)$, enhancing regularity and affecting horizon locations.

This construction interpolates smoothly between a regular core and the Schwarzschild geometry. Indeed, in the limiting case where both deformation parameters vanish, i.e., $(a,n) \to 0$, we recover the classical Schwarzschild solution.

The corresponding Hernandez-Misner-Sharp mass function for this model takes the form
\begin{equation}
M_{\text{HMS}}(r) = \frac{1}{2\sqrt{r^2 + a^2}} \left[ a^2 + \frac{2^{n+1} m r}{\pi^n} \arctan^n \left( \frac{r}{a} \right) \right] \label{massmod4}.
\end{equation}

For odd values of the exponent $n$, the HMS mass is guaranteed to remain positive throughout the entire spacetime. Moreover, this mass function satisfies the following limiting behaviors:
\begin{equation}
\lim_{r \to 0} M_{\text{HMS}}(r) = \frac{a}{2}, \qquad \lim_{r \to \infty} M_{\text{HMS}}(r) = m.
\end{equation}

\begin{figure}[ht]
\centering
\includegraphics[height=4.5cm]{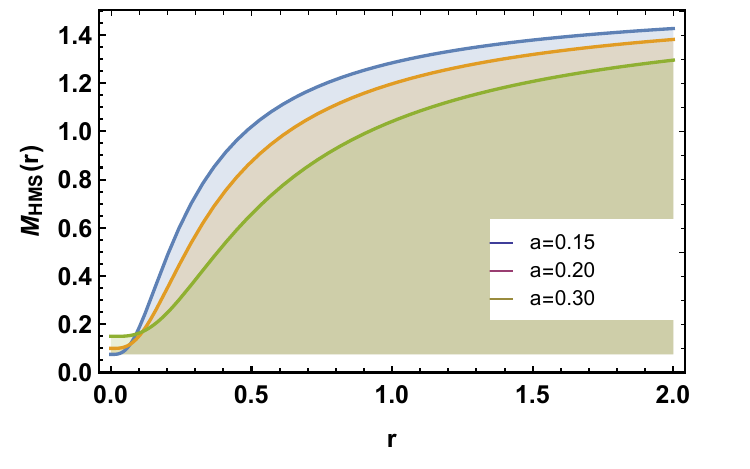}
\includegraphics[height=4.5cm]{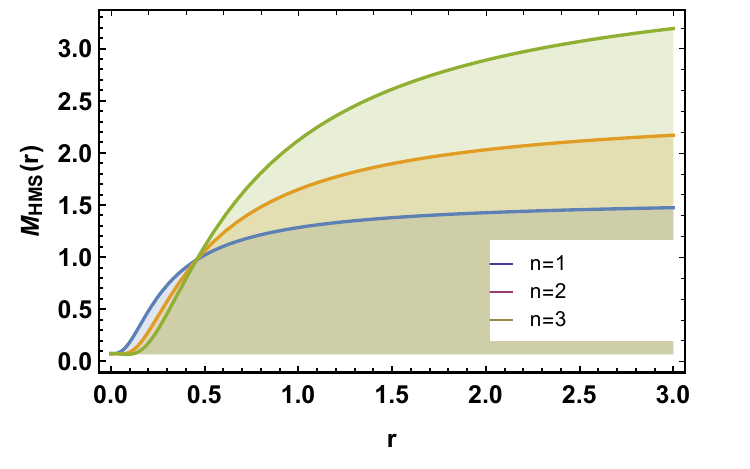}\\
\caption{Hernandez–Misner–Sharp quasi-local mass with $m=1$. (a) $n=1$. (b) $a=0.15$.}
\label{fig55}
\end{figure}

This plot (Figure (\ref{fig55})) shows $M_{\text{HMS}}(r)$ for the arctangent-based model with $m = 1$. (a) : fixed $n = 1$, and $a = 0.15, 0.20, 0.30$. (b) : fixed $a = 0.15$, and $n = 1, 2, 3$. Again, the quasi-local mass remains regular and grows monotonically, emphasizing the effectiveness of the modified mass function in eliminating singularities.

This choice of $M(r)$ offers a promising avenue for constructing geometries that avoid curvature singularities while maintaining a physically viable matter distribution.

\subsection{Model $M(r)=m\arctan^n(r/a)(2/\pi)^n$}

We introduce an alternative mass function designed to yield a positive energy density throughout the geometry. Specifically, consider
\begin{equation}
M(r) = m \left( \frac{2}{\pi} \right)^n \arctan^n\left( \frac{r}{a} \right),
\end{equation}
which leads to the following form for the lapse function:
\begin{equation}
f(r) = 1 - \frac{2 M(r)}{\Sigma(r)} = 1 - \frac{2m}{\sqrt{r^2 + a^2}} \left( \frac{2}{\pi} \right)^n \arctan^n\left( \frac{r}{a} \right) \label{mod5}.
\end{equation}

\begin{figure}[ht]
\centering
\includegraphics[height=4.5cm]{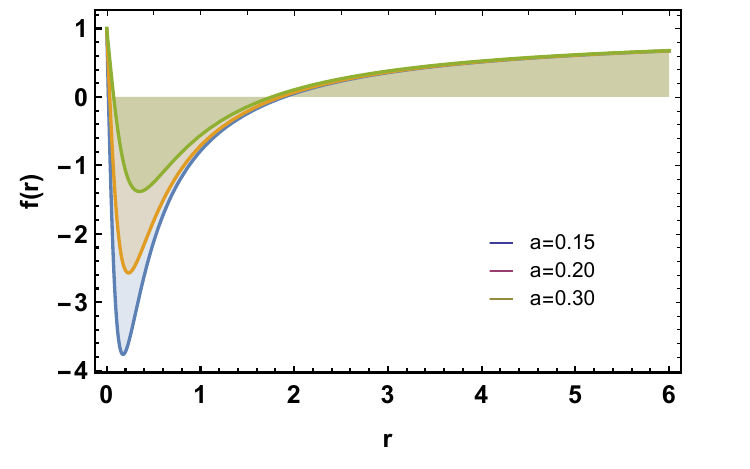}
\includegraphics[height=4.5cm]{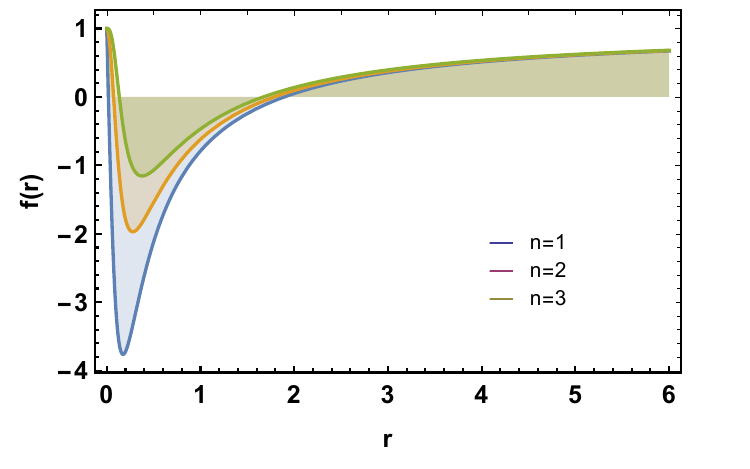}\\
\caption{Metric function $f(r)$ with $m=1$. (a) $n=1$. (b) $a=0.15$.}
\label{fig6}
\end{figure}

This figure (\ref{fig6}) presents $f(r)$ for the simpler arctangent model $M(r) = m \left(\frac{2}{\pi}\right)^n \arctan^n(r/a)$, with $m = 1$. (a) : fixed $n = 1$, and $a = 0.15, 0.20, 0.30$. (b) : fixed $a = 0.15$, and $n = 1, 2, 3$. Aslo, the behavior of $f(r)$ remains smooth and well-behaved, reinforcing the regular structure of the geometry.

This setup reduces to the Simpson-Visser geometry in the limit $n \to 0$, while taking $(a,n) \to 0$ recovers the Schwarzschild solution. Furthermore, the Kretschmann scalar remains finite for all $r$, indicating a regular geometry with no curvature singularities.

The corresponding Hernandez-Misner-Sharp mass function takes the form
\begin{equation}
M_{\text{HMS}}(r) = \frac{a^2}{2 \sqrt{r^2 + a^2}} + \frac{m \left( \frac{2}{\pi} \right)^n r^2 \arctan^n\left( \frac{r}{a} \right)}{r^2 + a^2} \label{massmod5}.
\end{equation}

For even values of $n$, this expression guarantees a positive mass function over the entire radial domain. The limiting behavior of the mass function is given by
\begin{equation}
\lim_{r \to 0} M_{\text{HMS}}(r) = \frac{a}{2}, \qquad \lim_{r \to \infty} M_{\text{HMS}}(r) = m,
\end{equation}
showing smooth interpolation between a de Sitter-like core and an asymptotically flat exterior.

\begin{figure}[ht]
\centering
\includegraphics[height=4.5cm]{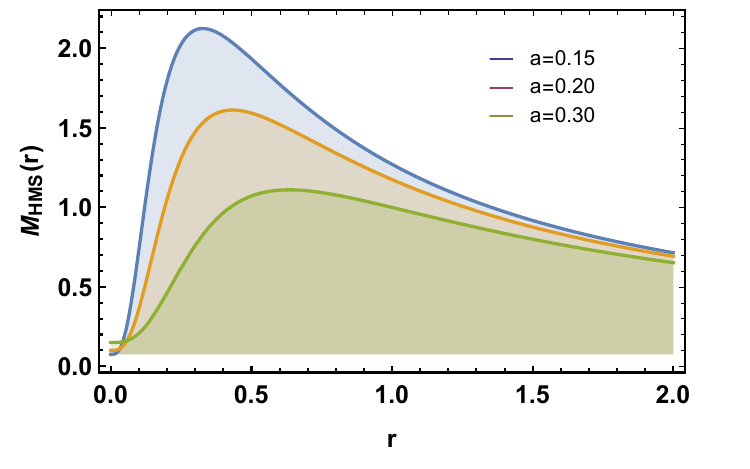}
\includegraphics[height=4.5cm]{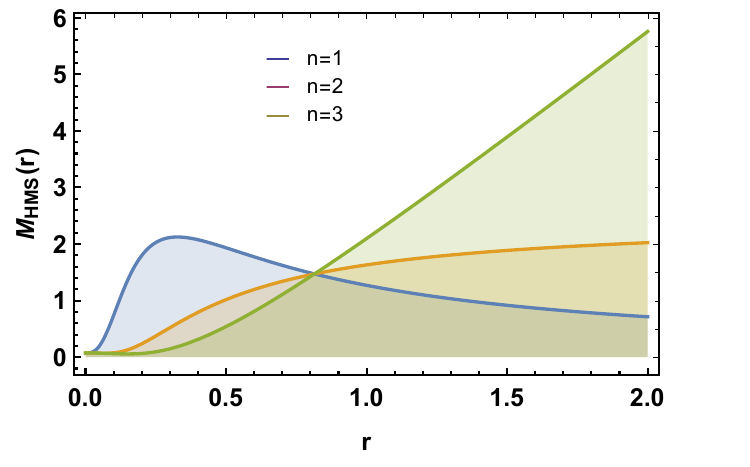}\\
\caption{Hernandez–Misner–Sharp quasi-local mass with $m=1$. (a) $n=1$. (b) $a=0.15$.}
\label{fig66}
\end{figure}

This figure (\ref{fig66}) shows $M_{\text{HMS}}(r)$ for the same arctangent model, with $m = 1$. (a) : $a = 0.15, 0.20, 0.30$, and fixed $n = 1$.    (b) : $n = 1, 2, 3$, and fixed $a = 0.15$. This model preserves the expected mass behavior, growing from $a/2$ to $m$ as $r \to \infty$.

This construction offers a regular and physically admissible black-bounce geometry, generalizing the Simpson-Visser model while ensuring improved energy conditions in specific regimes.

\section{Hawking temperature}

To investigate the temperature associated with black holes, we employ the Hamilton–Jacobi variant of the tunneling method~\cite{Srinivasan1998ty,Angheben2005rm,Kerner2006vu,Mitra2006qa,Akhmedov2006pg}. This semiclassical framework interprets Hawking radiation as a tunneling phenomenon, where quantum fluctuations near the event horizon lead to the spontaneous creation of particle-antiparticle pairs. One member of the pair, carrying negative energy, is absorbed by the black hole, reducing its mass, while the other escapes to infinity. The tunneling amplitude, in turn, encodes thermodynamic information such as the temperature. A key advantage of this approach is its reliance solely on the local structure of spacetime near the horizon, making it applicable to a wide class of static and dynamic geometries~\cite{Jiang2006,Kerner2007rr,Ma2014qma,Maluf2018lyu,Gomes2020kyj}.

Near the event horizon, the contributions from the angular part of the metric become negligible due to gravitational redshift. Therefore, the effective dynamics are governed by a two-dimensional metric of the form
\begin{align}
ds^2 = -A(r)dt^2 + A(r)^{-1}dr^2,
\end{align}
where the metric function $ A(r) $ vanishes at the event horizon. We consider a perturbative scalar field $ \phi $, with mass $ m $, evolving in this background, governed by the Klein–Gordon equation
\begin{align}
\hbar^2 g^{\mu\nu}\nabla_\mu \nabla_\nu \phi - m^2 \phi = 0.
\end{align}
Decomposing into spherical harmonics and focusing on the radial-temporal sector yields
\begin{align}\label{eq:kg_reduced}
-\partial_t^2 \phi + A(r)^2 \partial_r^2 \phi + \frac{1}{2} \partial_r A(r)^2 \partial_r \phi - \frac{m^2}{\hbar^2} A(r) \phi = 0.
\end{align}

Treating $ \phi $ as a semiclassical field, we apply the WKB approximation~\cite{Sakurai}, introducing the ansatz
\begin{align}
\phi(t,r) = \exp\left[\frac{1}{\hbar} \mathcal{T}(t,r)\right],
\end{align}
which, upon substitution into Eq.~\eqref{eq:kg_reduced} and retaining only the leading-order terms in $ \hbar $, yields
\begin{align}
(\partial_t \mathcal{T})^2 - A(r)^2 (\partial_r \mathcal{T})^2 - m^2 A(r) = 0.
\end{align}
Assuming a separable solution of the form
\begin{align}
\mathcal{T}(t,r) = -\omega t + W(r),
\end{align}
where $ \omega $ denotes the energy of the emitted particle, leads to the radial equation
\begin{align}
W(r) = \pm \int \frac{dr}{A(r)} \sqrt{\omega^2 - m^2 A(r)}.
\end{align}
The positive sign corresponds to the outgoing particle. Near the horizon, where $ A(r) $ vanishes, we expand the function linearly:
\begin{align}
A(r) \approx A'(r_H)(r - r_H).
\end{align}
Substituting into the integral gives
\begin{align}
W(r) = \int \frac{dr}{A'(r_H)} \frac{\sqrt{\omega^2 - m^2 A'(r_H)(r - r_H)}}{r - r_H}.
\end{align}

Evaluating this near-pole integral using the residue theorem, the imaginary contribution becomes
\begin{align}
\text{Im}[W(r_H)] = \frac{\pi \omega}{A'(r_H)}.
\end{align}
Consequently, the tunneling probability is given by
\begin{align}
\Gamma \sim \exp(-2\, \text{Im}[\mathcal{T}]) = \exp\left(-\frac{4\pi \omega}{A'(r_H)}\right),
\end{align}
which can be identified with the Boltzmann factor $ \Gamma \sim e^{-\beta \omega} $, where $ \beta = 1/T_{BH} $. Hence, the Hawking temperature associated with the black hole is
\begin{align}
T_{BH} = \frac{A'(r_H)}{4\pi}.
\end{align}

\begin{figure}[ht!]
\centering
\includegraphics[height=5cm]{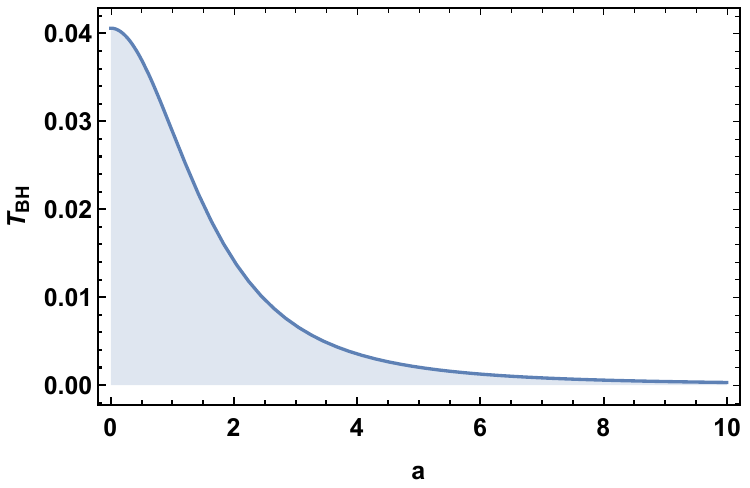}
\caption{Hawking Temperature with $m=1$.}
\label{fig7}
\end{figure}

This figure (\ref{fig7}) displays the Hawking temperature $T_{BH}$ as a function of the bounce parameter $a$ for $m = 1$. We notice that there is a maximum temperature. The interesting thing happens when we increase the value of $a$, where the temperature decreases until it reaches zero near $a\rightarrow\infty$. In other words, Fig.(\ref{fig7}) illustrates how the temperature decreases as $a$ increases, indicating reduced surface gravity. Furthermore, for $a=0$, we have the highest radiation index of the model. When we increase the value of $a$, the radiation felt at the event horizon decreases.

\begin{figure}[ht!]
\centering
\includegraphics[height=7cm]{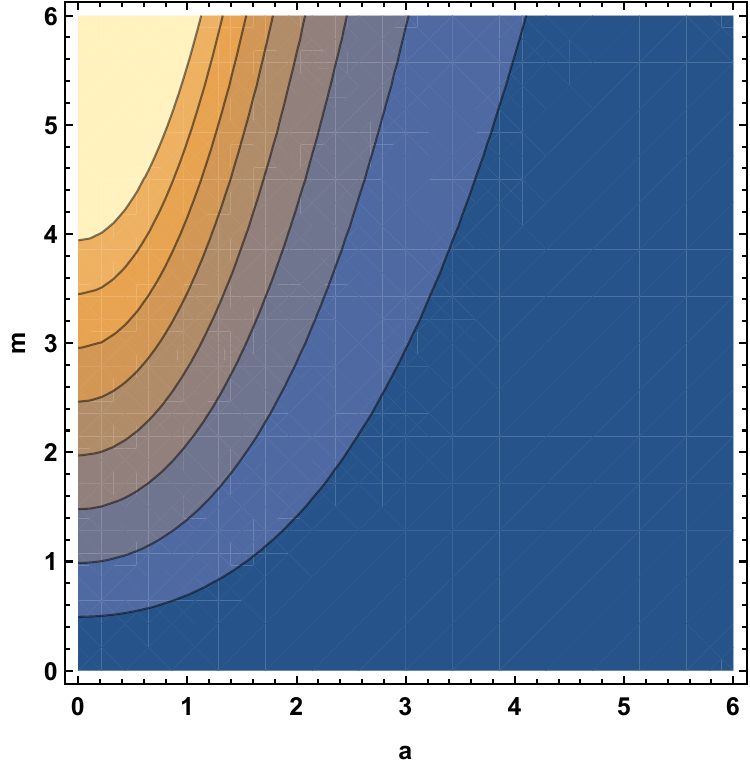}
\caption{Hawking Temperature varying the mass $m$.}
\label{fig777}
\end{figure}

For a more complete analysis, we plot in Fig.(\ref{fig777}) the behavior of the temperature when we vary the mass $m$ and the parameter $a$. Through this analysis we can identify which are the regions with the highest temperature, which are the yellow colored regions. The regions in dark blue are the regions where the temperature $T_{BH}=0$. Therefore, we can state that for values of $m\geq4$ the temperature assumes maximum values, but this is only valid for low values of $a<2$.

\subsection{Model $n=1$ and $k=2$}

\begin{figure}[ht]
\centering
\includegraphics[height=5cm]{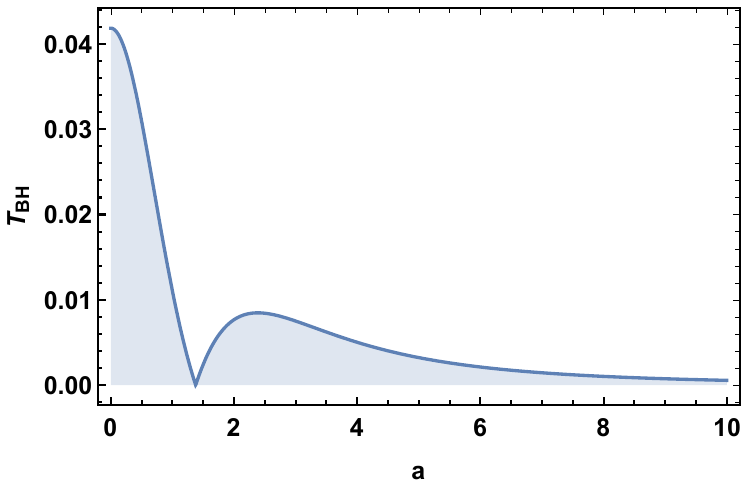}
\caption{Hawking Temperature with $m=1$.}
\label{fig8}
\end{figure}

The Fig. (\ref{fig8}) It shows a qualitatively similar temperature diminishes with increasing $a$, signifying more stable and colder black-bounce configurations. Also note that the temperature decreases in the interval $0<a<1.45$, and at $a\approx1.45$ we have a temperature equal to zero. The interesting thing happens after $a\approx1.45$, which presents an increase in temperature and then it tends to fall again $a\rightarrow\infty$. This indicates that when $a=0$, we have the configuration with the highest delay. However, this radiation decays rapidly, leading to a zero temperature at $a\approx1.45$. After that, we have a second temperature peak at $a\approx2.3$, and we have low radiation for higher values of $a$, i.e., there is a more stable configuration when the temperature decreases, which does not necessarily need to be for high values of $a$.

\begin{figure}[ht]
\centering
\includegraphics[height=7cm]{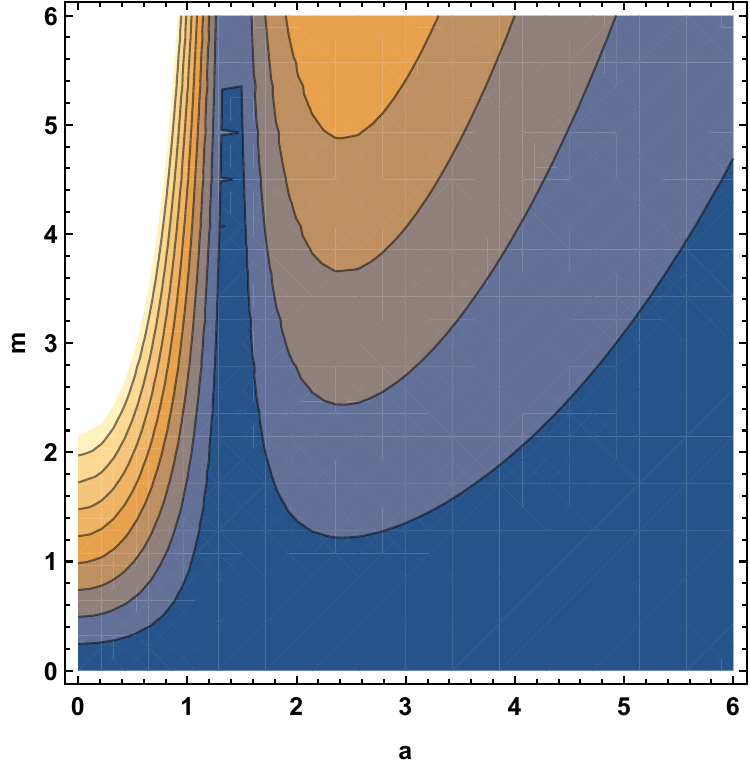}
\caption{Hawking Temperature varying the mass $m$.}
\label{fig888}
\end{figure}

In Fig.(\ref{fig888}) we plot the behavior of the temperature $T_{BH}$ for variations in mass $m$ and parameter $a$. The yellowish regions tending towards white are the regions of highest temperature. Note that we reach the highest temperature peak in the region $m\geq2.2$, but for low values of $a<1$. Furthermore, we observe the emergence of a second region of high temperatures for values between $2<a<3$, but for high values of $m>5$. Finally, we highlight that there are regions of zero temperature, which are highlighted by the dark blue color.

\subsection{Model $n=2$ and $k=0$}\label{SS:n=2+k=0}

\begin{figure}[ht]
\centering
\includegraphics[height=5cm]{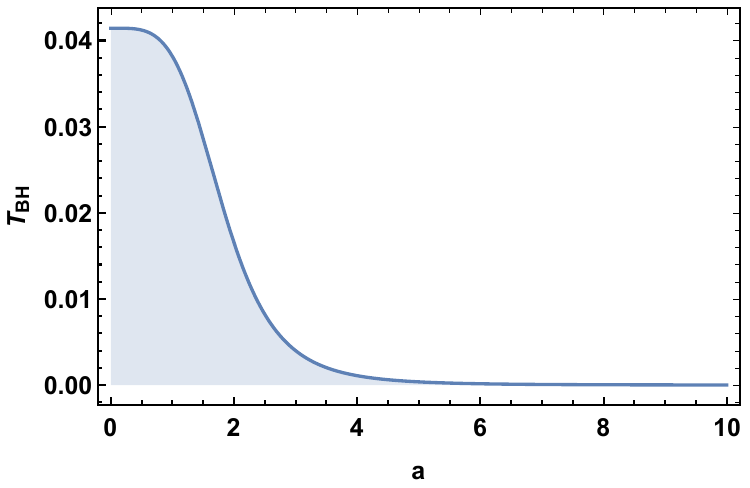}
\caption{Hawking Temperature with $m=1$.}
\label{fig9}
\end{figure}
In Figure (\ref{fig9}) we plot the Hawking temperature $T_{BH}$ for $m = 1$ versus $a$, with $m=1$. Here we can observe that there is a temperature peak between $0<a<0.4$, which indicates a stable configuration with no changes in radiation. However, after this interval the temperature tends to fall, this happens because of the low radiation. Aslo, the temperature declines with increasing $a$, again showing that a larger bounce leads to less energetic Hawking radiation.

\begin{figure}[ht]
\centering
\includegraphics[height=7cm]{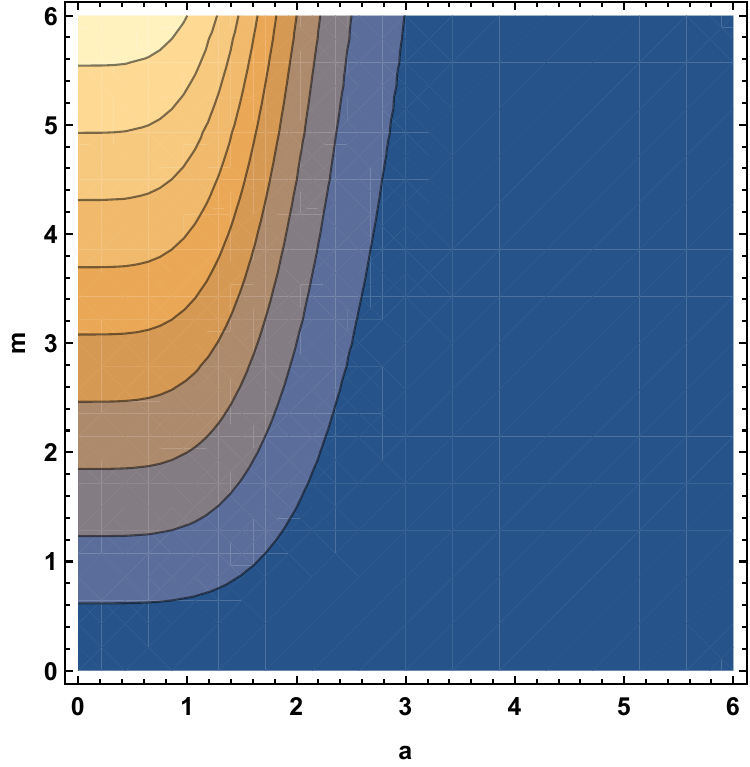}
\caption{Hawking Temperature varying the mass $m$.}
\label{fig999}
\end{figure}

Again, we analyze the behavior of the temperature $T_{BH}$ for variations of $m$ and $a$. For this, we plot in Fig.(\ref{fig999}) the temperature when varying the mass $m$ and the parameter $a$. The regions of highest temperature are the regions in yellow $m\geq5.5$, however, this region extends to low values of $a$, reaching $a\approx1$. The regions in dark blue are the regions where the temperature $T_{BH}=0$.

\subsection{Model $M(r)=m\cos^{2n}\left[r_0/\Sigma(r)\right]$}

\begin{figure}[ht]
\centering
\includegraphics[height=4.5cm]{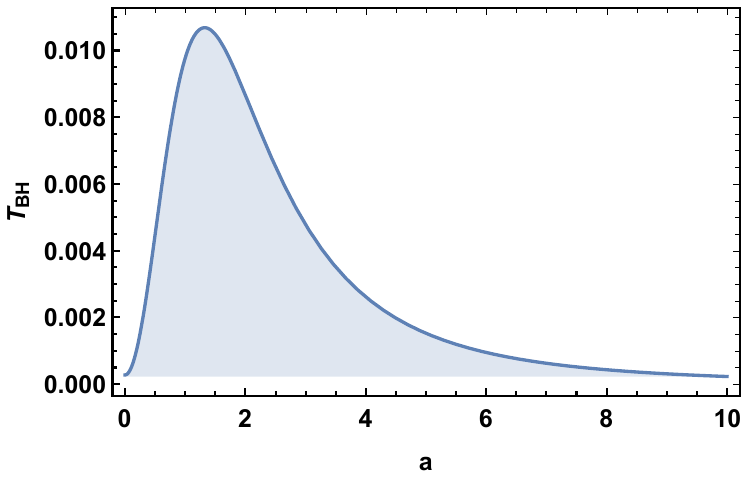}
\includegraphics[height=4.5cm]{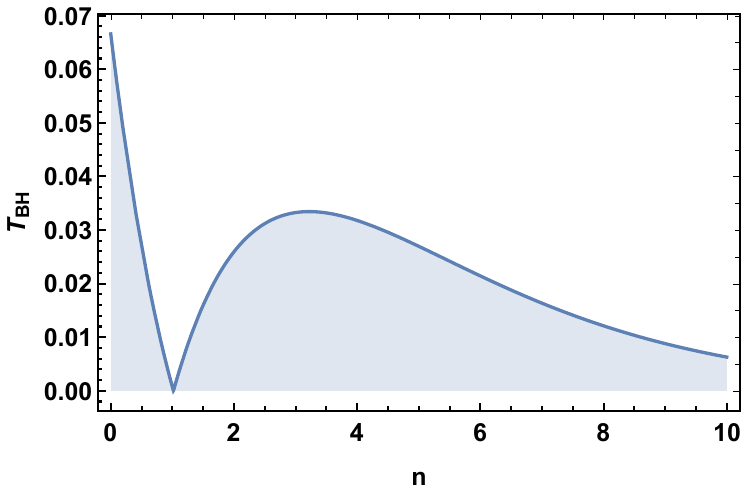}\\
\caption{Hawking Temperature with $m=1$. (a) $n=1$. (b) $a=0.15$.}
\label{fig10}
\end{figure}

This figure (\ref{fig10}) contains two plots: (a)  shows $T_{BH}$ versus $a$ for fixed $n = 1$.    (b)  shows $T_{BH}$ versus $n$ for fixed $a = 0.15$. In this case, these illustrate that both increasing $a$ and increasing $n$ suppress the Hawking temperature in this cosine-model geometry.

In Fig.(\ref{fig10}).a, we clearly notice that the temperature tends to zero when $a\rightarrow\infty$.
The same behavior is observed for $a=0$. Meanwhile, the temperature value is maximum at $a\approx1.86$. This is exactly what is expected for a Schwarzschild black hole and occurs due to the fact that the Schwarzschild temperature drops rapidly. From this, we can clearly see that the parameter $a$ directly influences the behavior of the thermodynamic properties of the model.

In Fig.(\ref{fig10}).b, we observe an anomalous behavior, where the temperature has its maximum value at $n=0$, regardless of the value of $a$. Furthermore, the temperature drops rapidly and reaches zero when $n\approx1$. After that, the temperature has a behavior expected of a Schwarzschild black hole, where $T_{BH}$ tends to zero when $n\rightarrow\infty$. Its second maximum point is $n\approx3.2$. This behavior is a clear reflection of the influence of the model's geometry. This indicates that the parameter $n$ will directly influence the behavior of the model's thermodynamic properties.

\begin{figure}[ht]
\centering
\includegraphics[height=7cm]{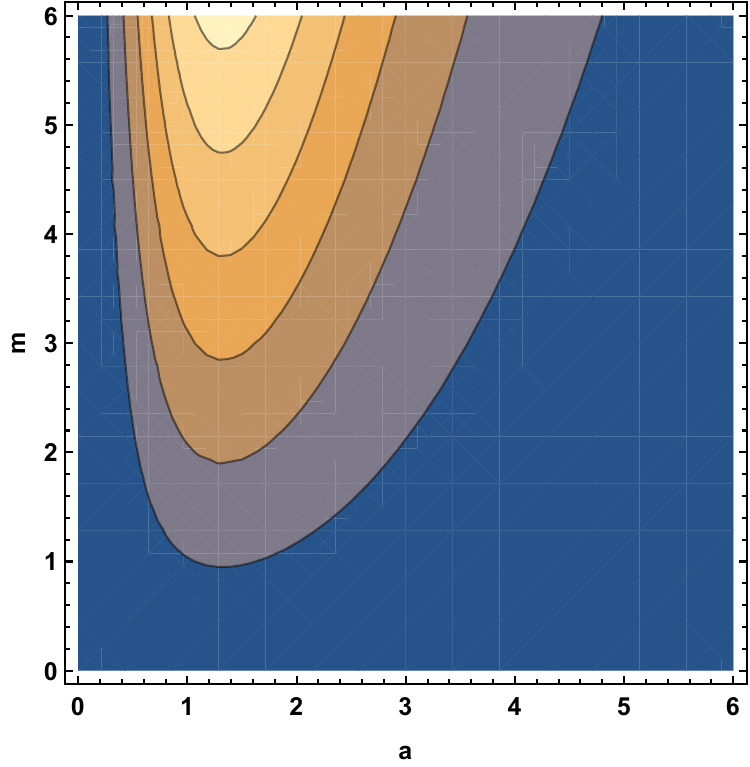}
\includegraphics[height=7cm]{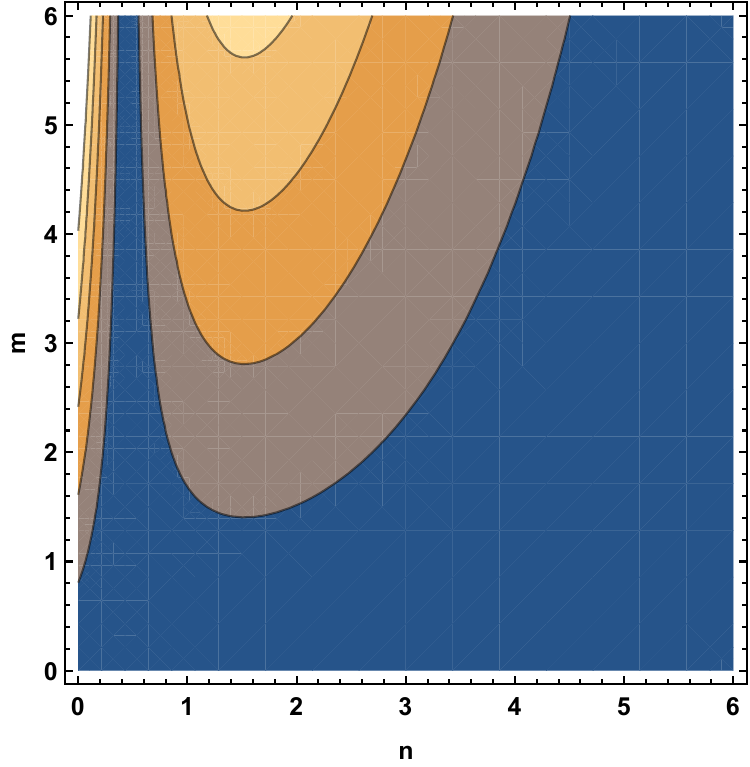}\\
\caption{Hawking Temperature varying the mass $m$. (a) $n=1$. (b) $a=0.15$.}
\label{fig1000}
\end{figure}

In Fig.(\ref{fig1000}) we plot the behavior of the temperature $T_{BH}$ for variations in mass $m$ and in the parameters $a$ and $n$. The yellowish regions tending towards white are the regions of highest temperature.
In Fig.(\ref{fig1000}).a, we note that the high-temperature region is at $1<a<1.6$ and $m>5.7$. At $a=0$, we have a region of zero temperature, which is represented by the dark blue regions. In Fig.(\ref{fig1000}).b,
Note that we reach the highest temperature peak in the region $m\geq4$, but for low values of $a<0.1$. In addition, we observe the emergence of a second high-temperature region for values between $1.2<a<1.9$, but for high values of $m>5.6$.

\subsection{Model $M(r)=m\arctan^n(r/a)\;(\Sigma/r) (2/\pi)^n$}

\begin{figure}[ht]
\centering
\includegraphics[height=4.5cm]{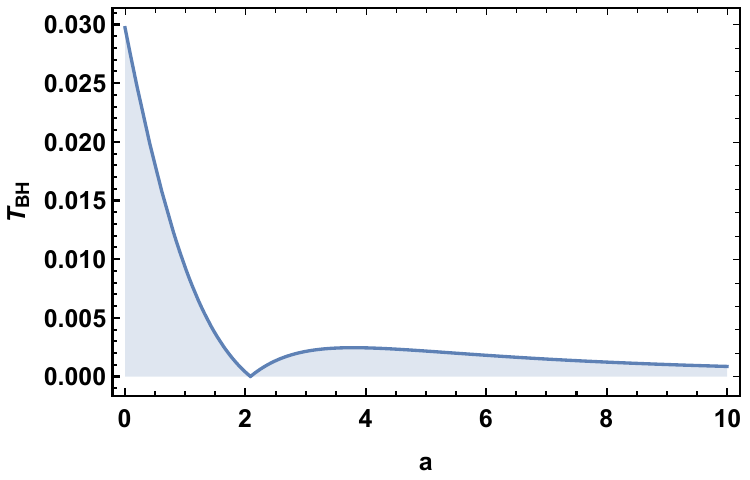}
\includegraphics[height=4.5cm]{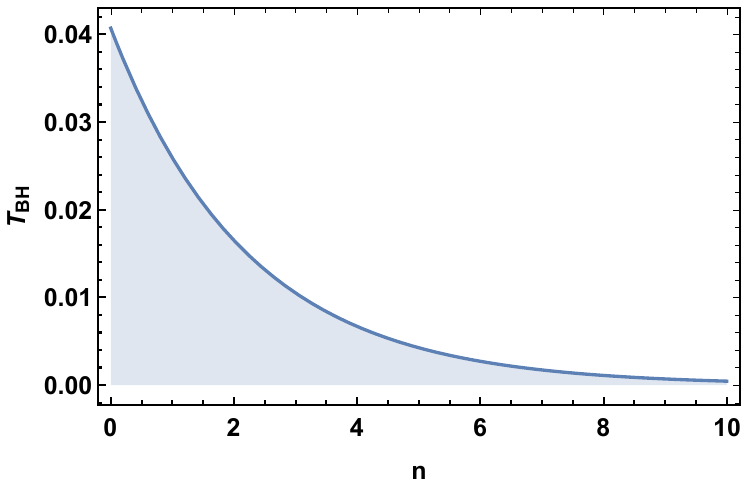}\\
\caption{Hawking Temperature with $m=1$. (a) $n=1$. (b) $a=0.15$.}
\label{fig11}
\end{figure}

For the arctangent–product model $M(r) = m \arctan^n(r/a) (\Sigma/r)(2/\pi)^n$, this figure (\ref{fig11}) show: (a) $T_{BH}$ vs. $a$ for fixed $n = 1$. (b) $T_{BH}$ vs. $n$ for fixed $a = 0.15$. It illustarte a steeper decline in temperature with increasing deformation parameters, emphasizing the strong suppression of radiation.

In Fig.(\ref{fig11}).a we observe a sharp drop in temperature for the first variations of $a$. When the temperature decreases to zero ($a\approx2.1$), the temperature increases slightly, reaching its second maximum at $a\approx3.6$ and then goes to zero for higher values of $a$. This indicates configurations with low radiation both for a very large value of $a$ and for a specific value of $a\approx2.1$. 

In Fig.(\ref{fig11}).b we observe an exponential drop as the value of $n$ increases. This shows how the geometry changes the thermodynamic properties, leading to low temperatures, which represent a configuration with greater stability.

\begin{figure}[ht]
\centering
\includegraphics[height=7cm]{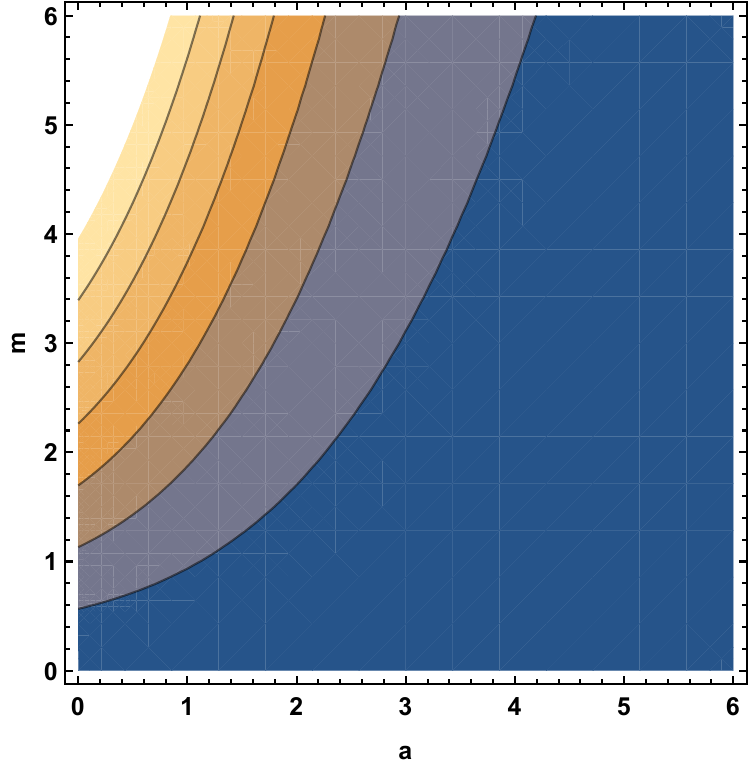}
\includegraphics[height=7cm]{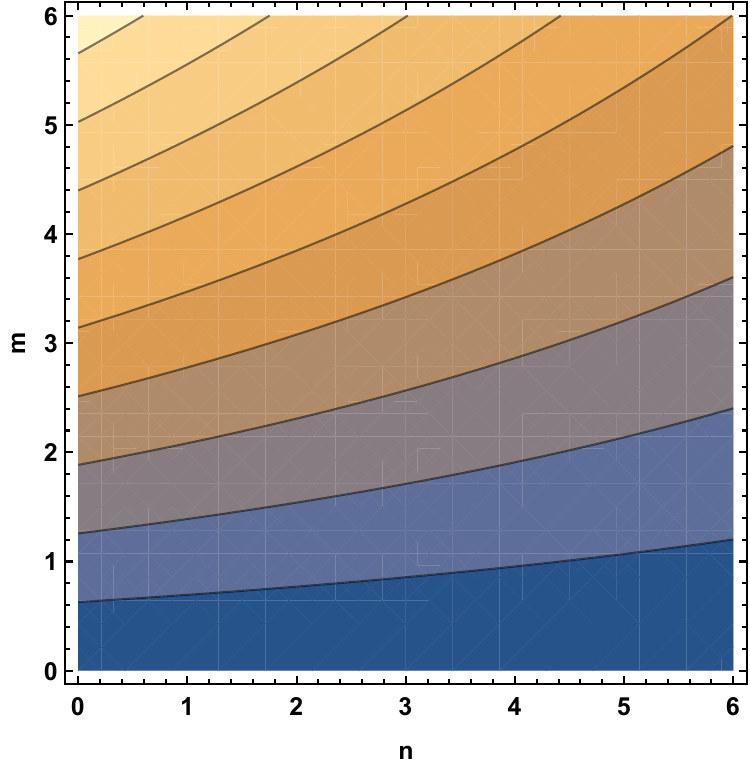}\\
\caption{Hawking Temperature varying the mass $m$. (a) $n=1$. (b) $a=0.15$.}
\label{fig1111}
\end{figure}

We plot the temperature behavior by varying the mass values $m$ and the parameters $a$ and $n$, in Fig.(\ref{fig1111}). We can observe that the high temperature values are located in the regions where $m\geq4$ and $a<0.9$. Furthermore, we can say that the high temperatures are reached only for $n\rightarrow0$. We also note that the dark blue regions are the zero temperature regions, i.e., the coldest and most stable regions.

\subsection{Model $M(r)=m\arctan^n(r/a)(2/\pi)^n$}

\begin{figure}[ht]
\centering
\includegraphics[height=4.5cm]{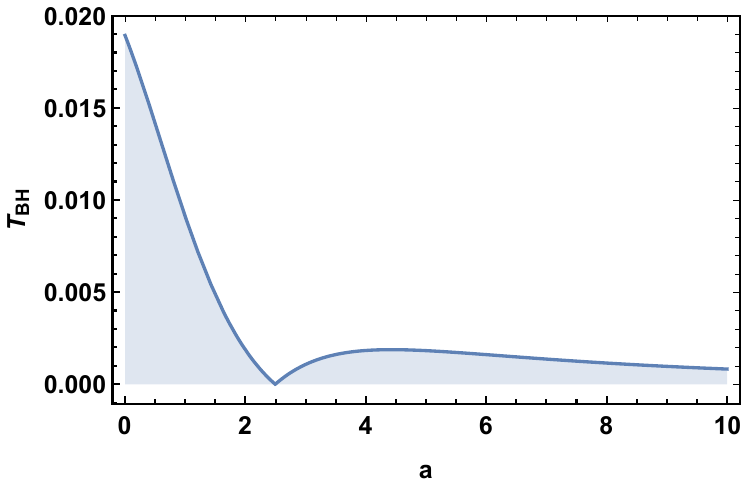}
\includegraphics[height=4.5cm]{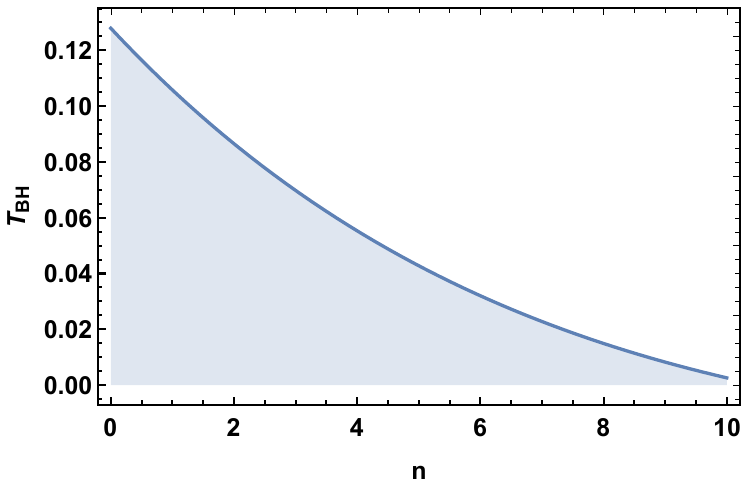}\\
\caption{Hawking Temperature with $m=1$. (a) $n=1$. (b) $a=0.15$.}
\label{fig12}
\end{figure}

This final figure (\ref{fig12}) represents the Hawking temperature for the pure arctangent model: (a) $T_{BH}$ as a function of $a$ for fixed $n = 1$.  (b) $T_{BH}$ versus $n$ for $a = 0.15$. Much like the previous models, the results again confirm that higher $a$ or $n$ correlates with lower black hole temperatures, reflecting softer geometries and more stable thermodynamic states. 

In Fig.(\ref{fig12}).a we observe that there is a point of minimum temperature different from $a\rightarrow\infty$, which is located at $a\approx2.45$. After this point of low radiation, the temperature increases and reaches its second maximum at $a\approx4.2$, which represents a point of maximum radiation. Afterwards, for high values of $a$ the temperature only tries to decrease reaching zero, i.e., reaching a stability configuration. As for Fig.(\ref{fig12}).b, we observe an exponential decay of the temperature when we increase the value of $n$, reaching zero temperature quickly.

\begin{figure}[ht]
\centering
\includegraphics[height=7cm]{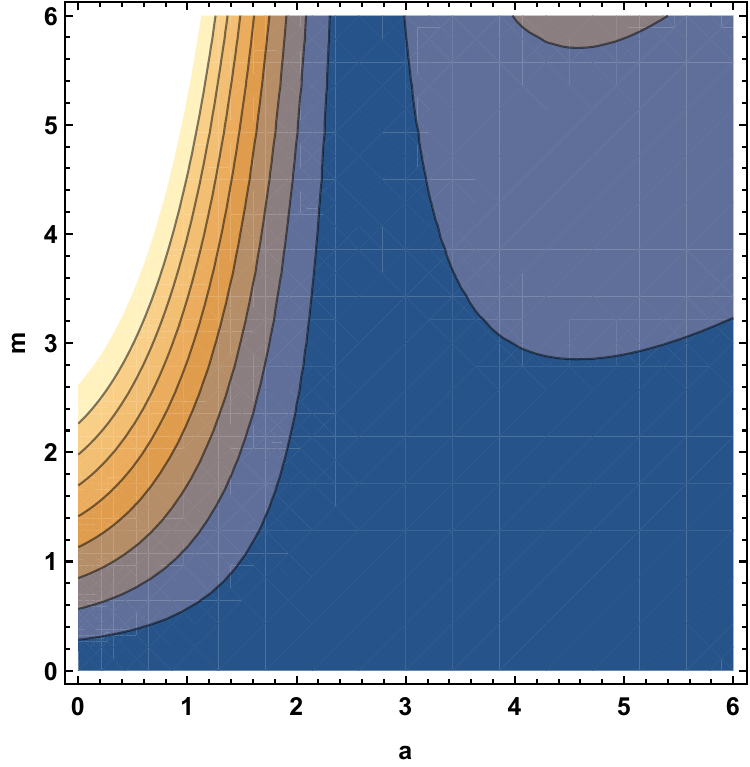}
\includegraphics[height=7cm]{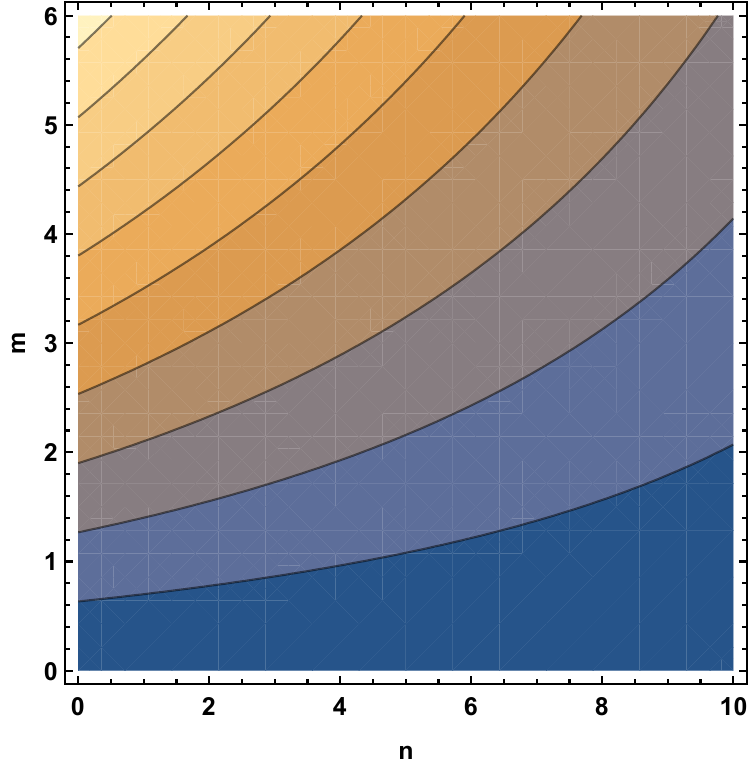}\\
\caption{Hawking Temperature varying the mass $m$. (a) $n=1$. (b) $a=0.15$.}
\label{fig1212}
\end{figure}

In Fig.(\ref{fig1212}), we present the temperature profile as a function of the mass parameters $m$, $a$ and $n$. It can be seen that the highest temperatures are concentrated in the regions where $m \geq 2.8$ and $a < 1.1$. Another region appears around $4<a<5.5$ with $m>5.8$, but with a peak temperature not as high as the first.
In addition, such high temperatures occur predominantly at the limit $n \rightarrow 0$. On the other hand, the dark blue regions correspond to zero temperatures, indicating states of greater thermal stability and, therefore, colder regions of the system.

\section{Conclusion}\label{sec00}

In this work, we constructed and analyzed a broad family of regular black-bounce spacetimes, extending the original Simpson–Visser geometry by introducing generalized mass functions and areal radii. By considering various functional forms for $M(r)$, including power-law, trigonometric, and arctangent-based profiles, we demonstrated the feasibility of generating nonsingular geometries that interpolate smoothly between Schwarzschild-like asymptotics and regular de Sitter-like cores.

For each model, we derived and examined the corresponding lapse functions $f(r)$ and quasi-local mass profiles $M_{\text{HMS}}(r)$, identifying the conditions under which these geometries remain regular and physically admissible. In particular, we showed that the bounce parameter $a$ plays a central role in regulating the depth of the gravitational potential and the behavior near the core, while the deformation parameters $n$ and $k$ control the structure of the horizon and energy distributions. Importantly, all proposed configurations satisfy the key criteria for regularity: the finiteness of curvature invariants, smoothness of the metric functions, and positivity of the quasi-local mass. Furthermore, we highlighted how specific parameter choices can improve the energy conditions in regions outside the horizon, making some of these solutions promising candidates for describing nonsingular compact objects or quantum-corrected black holes.

Through the Hamilton–Jacobi variant of the tunneling method, we systematically explored the Hawking temperature of various regular black hole models characterized by distinct mass functions $M(r)$. Each configuration introduces parameters such as the bounce parameter $a$, deformation index $n$, and black hole mass $m$, all of which play a central role in modifying the thermal behavior of the spacetime. In all models considered, we observed that increasing the deformation parameters $a$ and $n$ generally leads to a suppression of the Hawking temperature. This suggests a natural mechanism by which the geometry of the black hole interior, particularly the degree of regularity or deviation from the classical singularity, directly impacts the efficiency of Hawking radiation. Such behavior supports the interpretation that larger deformations correspond to softer spacetime geometries, reducing surface gravity and thus the radiation flux. Another common feature is the existence of critical points in the parameter space, values of $a$ or $n$ where the temperature either vanishes or peaks. These extrema indicate transitions between radiative and non-radiative regimes and may be interpreted as phase-like transitions in the thermodynamic portrait of regular black holes. In particular, the appearance of zero-temperature configurations (dark blue regions in the plots) marks the onset of extremal or quasi-stable states where quantum radiation is significantly suppressed or ceases altogether. The influence of the mass $m$ is also noteworthy. While increasing $m$ can enhance the temperature for certain parameter regions, especially at small $a$, we find that this effect saturates or even reverses in highly deformed geometries. This nontrivial interplay underscores the fact that regularity effects cannot be captured merely by rescaling Schwarzschild-like behaviors, they must be understood within the full context of modified geometries.

From a broader perspective, these results reinforce the idea that regular black holes, those avoiding singularities via internal geometric deformations, admit richer thermodynamic structures than their classical counterparts. The existence of suppressed or vanishing temperatures opens avenues for stable remnant scenarios and suggests potential endpoints for black hole evaporation that evade the classical information loss paradox.
Future work may focus on dynamical extensions of these geometries, including rotating generalizations, stability analyses, or embedding them in modified gravity theories. The thermodynamic properties and causal structure of these generalized black-bounce spacetimes also deserve further investigation to assess their viability in realistic astrophysical scenarios.


\section*{Data Availability}

No new data were generated or analyzed in this study.

\section*{Conflict of Interests}

Authors declares there is no conflict of interests.

\section*{Funding Statement}

No fund has received for this study.


\end{document}